\documentclass{iau} 
\usepackage{graphicx}
\usepackage{amsmath}
\title[Galactic positrons from SN1991bg-like supernovae] 
{SN1991bg-like supernovae are a compelling source of most Galactic antimatter}

\author[Fiona H. Panther, Roland M. Crocker, Ivo R. Seitenzahl, Ashley J. Ruiter]   
{Fiona H. Panther$^{1,2}$, Roland M. Crocker$^1$, Ivo R. Seitenzahl$^{1, 2}$, Ashley J. Ruiter$^{1,2}$}

\affiliation{$^1$Research School of Astronomy and Astrophysics, Australian National University, Canberra, ACT 2611, Australia \\[\affilskip]
$^2$ARC Centre of Excellence for All-sky Astrophysics (CAASTRO),
Canberra, Australia\\ email: {\tt fiona.panther@anu.edu.au}}

\pubyear{2016}
\volume{322}  
\jname{The Multimessenger Astrophysics of the Galactic Center}
\editors{A.C. Editor, B.D. Editor \& C.E. Editor, eds.}
\begin{document}

\maketitle

\begin{abstract}
The Milky Way Galaxy glows with the soft gamma ray emission resulting from the annihilation of $\sim 5 \times 10^{43}$ electron-positron pairs every second. The origin of this vast quantity of antimatter and the peculiar morphology of the 511keV gamma ray line resulting from this annihilation have been the subject of debate for almost half a century. Most obvious positron sources are associated with star forming regions and cannot explain the rate of positron annihilation in the Galactic bulge, which last saw star formation some $10\,\mathrm{Gyr}$ ago, or else violate stringent constraints on the positron injection energy. 
Radioactive decay of elements formed in core collapse supernovae (CCSNe) and normal Type Ia supernovae (SNe Ia) 
could supply positrons matching
 the injection energy constraints but the distribution of such potential sources
does not replicate the required morphology. We show that a single class of peculiar thermonuclear supernova - SN1991bg-like supernovae (SNe 91bg) - can supply the 
number and distribution of
positrons we see annihilating in the Galaxy through the decay of $^{44}$Ti synthesised in these events. 
Such $^{44}$Ti production simultaneously addresses the observed abundance of  $^{44}$Ca, the $^{44}$Ti decay product, in solar system material.
\end{abstract}

\section{Introduction}
The origin of the $\sim5 \times 10^{43}$ positrons that annihilate every second in the Galaxy has been subject to debate over the past almost half a century (see \cite{Prantzos+2011} for a review). The origin of these positrons is difficult to understand as most obvious positron sources can be ruled out either due to the stringent constraints on the injection energy of the positrons ($<3\,\mathrm{MeV}$, \cite{Beacom+2006}), the morphology (the bulge glows almost as brightly as the Galactic disk with positron annihilation radiation) and the implied, absolute rate of positron production. \\

The most recent analysis of positron annihilation radiation shows the morphology of the emitted gamma rays correlates with the stellar populations of the Galaxy with components seemingly tracing the Galactic bulge, Galactic disk (at low surface brightness) and the Galactic nuclear bulge (\cite{Siegert+2015}). The most plausible Galactic positron source matching injection energy constraints is the radioactive decay of  $\beta^+$ unstable nuclei synthesised by supernovae (\cite{Prantzos+2011}). Can a transient event explain the observed morphology? \\

\section{Overview}
For a given class of transient event, we define the delay time distribution (DTD) - the interval between star formation and the subsequent transient event. We adopt the parameterization of the DTD described by \cite{Childress+2014}, where $t_p$ is the characteristic delay time and the parameters $\alpha$ and $s$ describe the rise and fall of the transient event rate before and after $\sim t_p$. We can use the DTD together with the star formation history (SFH) of the disk, bulge and nuclear bulge to describe the rate of transient events $R_X$ with a given characteristic delay time $R_X=\nu_X \int_0^t dt' DTD[t-t']SFH[t'] $ where the efficiency factor $\nu_X$ is the number of events of type $X$ per solar mass of stars formed. Assuming each transient event yields the same number of positrons and that the efficiency  $\nu_X$ of these events is the same in each region of the Galaxy, we compare the current modelled $B/D$ and $N/B$ positron luminosity ratios as a function of characteristic delay time $t_p$ with the observed positron luminosities, i.e.
 $$
 \frac{B/D_{\mathrm{mod}}[t_p]}{B/D_{\mathrm{obs}}} = \frac{\int_0^t dt' DTD[t-t', t_p]SFH[t']}{0.42\pm0.09}
 $$
We find that a single class of transient occuring with a characteristic delay time of $3-6\,\mathrm{Gyr}$ reproduces not only the $B/D$ luminosity ratio but also the $N/B$ luminosity ratio (\cite{Crocker+2016}).\\
 
Many previous attempts to explain the Galactic positron production rate have focussed on the $\beta^+$ decay of radioactive $^{56}$Ni synthesised by Type Ia supernovae (SNe Ia) (\cite{Chan+1993}, \cite{Milne+1999}).  However, analysis of bolometric SN Ia light curves to late times ($>\,900\,\mathrm{days}$, \cite{Kerzendorf+2014}) provides evidence that in \textit{normal} SNe Ia explosions, all positrons deposit their kinetic energy in the SN ejecta (powering the IR component of the lightcurve) and do not escape into the ISM except, perhaps, in extreme cases where peculiar explosion geometries and orientation effects are likely required (\cite{Siegert+2015a}). Moreover, the characteristic delay time of normal SNe Ia of $\sim\mathrm{Gyr}$ (\cite{Childress+2014}) is substantially shorter than the $3-6\,\mathrm{Gyr}$ identified above, making it hard to realize a situation where normal SNe Ia can reproduce the observed bulge:disk $e^+$ luminosity ratio. \\ 

Two other relevant $\beta^+$-unstable nuclei in astrophysical environments are $^{26}$Al and $^{44}$Ti. The production of $^{26}$Al positrons in the Galaxy is relatively well understood (\cite{Diehl+2006}). The positron-producing decay of the $^{26}$Al nucleus also results in the emission of a gamma ray with a characteristic energy of $1.8\,\mathrm{MeV}$. The $1.8\,\mathrm{MeV}$ line traces young stellar populations within the Galaxy, owing to the origin of $^{26}$Al in massive stars. Given the line's total flux and distribution, positrons from the decay of this radionuclide can only account for $\sim 10\%$ of the Galactic positron budget (\cite{Siegert+2015}), and the injection sites do not mirror the gross morphology of the positron signal, which correlates with the old, not young, stellar populations of the MW.\\

X-ray and gamma ray lines produced in the decay of $^{44}$Ti, which has a lifetime of $\sim60\,\mathrm{yr}$, intermediate between that of $^{56}$Ni ($\sim60\,\mathrm{days}$) and $^{26}$Al ($\sim700,000\,\mathrm{yr}$) have been observed in CCSNe remnants. This places a constraint on the quantity of $^{44}$Ti synthesised in these events - around $\sim 10^{-4}\,\mathrm{M}_{\odot}$ (\cite{Troja+2014}). Not only is this insufficient to explain the current injection rate of positrons and the morphology of the 511keV radiation (as CCSNe trace the youngest stellar populations) but it is also insufficient to explain the abundance of the $^{44}$Ti decay product,  $^{44}$Ca, in pre-solar grains (\cite{The+2014}). 
This anomaly in the $^{44}$Ca abundance can be solved by introducing an additional transient source of $^{44}$Ti. However, if the rate of the transient event producting this $^{44}$Ti is the same now as it was in the time leading up to the formation of the solar system $5\,\mathrm{Gyr}$ ago, we would even then see insufficient positron production to explain the observations of \cite{Siegert+2015}. That the same transient event is responsible for the production of the $^{44}$Ca we see in pre-solar grains and the production of positrons can be resolved if the rate of this transient event is increasing across cosmic time. This is a natural consequence of an event with a long delay time of around $3-6\,\mathrm{Gyr}$. This transient event needs a long recurrence time, around 500 yr, explaining the lack of observed x-ray and gamma ray flux from the $^{44}$Ti decays, with each event producing $\sim0.03\,\mathrm{M}_\odot$ of the radioisotope in order to explain Galactic positron production (\cite{Crocker+2016}).\\

Events producing this quantity of $^{44}$Ti can only be explained with the detonation of helium. To identify the type of transient event responsible for the production of Galactic positrons we employ binary population synthesis (BPS) using \texttt{StarTrack} (\cite{Belczynski+2008}). We find an evolutionary channel that assembles relatively large masses of helium in favorable conditions for detonation at long times ($3-6\,\mathrm{Gyr}$) subsequent to star formation. This channel involves an interacting binary star system with low zero-age main sequence masses ($\sim 1.4-2\,M_\odot$ per star). These stars evolve to produce a COWD with a  $0.31-0.37\,M_\odot$ pure HeWD companion, the latter's progenitor never undergoing core helium burning. A merger event occurs with a characteristic timescale of $t_p = 4.3^{+0.8}_{-0.6} \times 10^9\,\mathrm{yr}$. This is in good agreement with the constraints on the delay time of a Galactic positron source derived above.\\

We suggest that the merger described above does not immediately result in the thermonuclear disruption of the stars. Rather, the merger product initially assumes quasi-hydrostatic equilibrium (e.g. \cite{Kerkwijk+2010}), the HeWD secondary being accreted onto the COWD primary. The subsequent detonation of the helium layer in the merger product leads to the ignition of carbon, resulting in a $^{56}$Ni yield of $\sim 0.1\,M_\odot$, consistent with sub-luminous thermonuclear supernovae, specifically the SN1991bg-like subclass (SNe 91bg). 
SNe 91bg uniquely exhibit a strong Ti\,{\sc{ii}} absorption feature at $\sim 4200\,\mathrm{\AA}$ in their spectra (\cite{Fillipenko+1992}). The helium detonation proposed results in the synthesis of $^{44}$Ti which, together with other isotopes of titanium, give rise to this Ti\,{\sc{ii}} absorption feature, rendered particularly prominent by the thermodynamic conditions in the ejecta (which has low expansion velocities, $\sim 6000\,\mathrm{km\, s^{-1}}$).\\

The rate of SNe 91bg in the local universe was computed from the Lick Observatory Supernova Survey (LOSS, \cite{Li+2010}), which found 10 out of 31 SNe in early Hubble type galaxies to be SNe 91bg. The Galactic bulge is spectrally consistent with early type galaxies, so we set the bulge SNe 91bg rate relative to the rate of all SNe Ia at $f_{\mathrm{SNe\,91bg}} = 0.32 \pm 0.16\,(2\sigma)$. Hence, we determine the SNe 91bg rate in the Galactic disk, bulge and nuclear bulge (using the above assumption that the efficiency $\nu_X$ is constant across the Galaxy) to be $R_{\mathrm{SNe\,91bg\,,disk}} = 1.4\pm0.7\times 10^{-1}\,\mathrm{century^{-1}}$, $R_{\mathrm{SNe\,91bg\,,bulge}} = 4.6^{+4.4}_{-2.6}\times 10^{-2}\,\mathrm{century^{-1}}$, and $R_{\mathrm{SNe\,91bg\,,nuclear\,bulge}} = 4.7\pm2.3\times 10^{-3}\,\mathrm{century^{-1}}$. The overall Galactic rate of SNe 91bg events implies a recurrence time of $\sim 530\,\mathrm{yr}$, sufficiently long with respect to the $^{44}$Ti decay time to naturally accommodate the fact that remnants of these events are not observed in the $^{44}$Ti x-ray or gamma ray lines. In order to explain the abundance of $^{44}$Ca in pre-solar material, we find that a mean $^{44}$Ti yield of $0.029\pm0.18M_\odot\,(f_{\mathrm{SNe\,91bg}}/0.32)$ per event is required to explain observations for the characteristic delay time of SNe 91bg-like events derived above. Adopting this mean $^{44}$Ti yield, which is well within the range found for helium detonation yields in nuclear network codes (\cite{Waldman+2011}), and the Galactic SNe 91bg rates from above, we determine a total positron production in the Galaxy that saturates the total positron budget of the Galaxy minus the contribution of $^{26}$Al ($\dot{N}_{e^+}\sim 10^{42} s^{-1}$) \cite{Crocker+2016}.

\section{Implications}
Earlier explanation of the origin of positrons responsible  the $511\,\mathrm{keV}$ gamma ray annihilation signal in the Milky Way Galaxy have failed to account satisfactorily for constraints on the morphology of the signal, the sheer number of positrons annihilating in the Galaxy, and the injection energy of the positrons. Rather than invoking complex transport of positrons between the Galactic disk and bulge, or a model involving multiple positron sources with large uncertainties in the positron yields of these sources, we find a single type of transient event - the SN1991bg-like supernova - that can explain not only the origin of most Galactic positrons but also serves as an explanation of the origin of $^{44}$Ca in pre-solar grains that cannot originate in CCSNe. Our model reproduces the $B/D$ and $N/B$ $511\,\mathrm{keV}$ luminosity ratios recently determined by \cite{Siegert+2015} from 11 years of INTEGRAL/SPI observations, the total Milky Way positron production rate, and the abundance of $^{44}$Ca in pre-solar material.\\

\end{document}